\documentclass[prd,twocolumn,showpacs,preprintnumbers,amsmath,amssymb]{revtex4}
\usepackage{graphicx}
\begin{document}

\setlength{\topmargin}{-0.25in}
\title{High-accuracy critical exponents of $O(N)$ hierarchical sigma models}
\author{J. J. Godina}
\affiliation{Dep. de Fisica\\ CINVESTAV-IPN\\ Ap. Post. 14-740, Mexico, D.F. 07000}
\author{L. Li} 
\author{Y. Meurice}
\email[]{yannick-meurice@uiowa.edu}
\affiliation{Department of Physics and Astronomy\\ The University of Iowa\\
Iowa City, Iowa 52242, USA }
\author{M. B. Oktay}
\affiliation{School of Mathematics\\Trinity College\\
Dublin, Ireland
}
\date{\today}
\begin{abstract}
We perform high-accuracy calculations of the critical exponent $\gamma$ and its subleading 
exponent for 
the $3D$ $O(N)$ Dyson's hierarchical model for $N$ up to 20. 
We calculate the critical temperatures for the nonlinear sigma model measure $\delta(\vec{\phi} . \vec{\phi}-1)$.
We discuss the possibility of extracting the first coefficients of the $1/N$ expansion 
from our numerical data. We show that the leading and subleading exponents agree
with Polchinski equation  and the equivalent Litim equation, in the local potential approximation, with at least 4 significant digits.

\end{abstract}
\maketitle

The large $N$ limit and the $1/N$ expansion \cite{stanley68,ma73,thooft74} appear prominently in recent developments in particle physics, condensed matter and string theory \cite{teper05,narayanan05,moshe03,aharony99}. 
For sigma models, the basic gap equation can be obtained by using the method of steepest descent for the functional integral \cite{stanley68,david85}. For $N$ large and negative, the maxima of the action dominate instead of the minima and the radius of convergence of the $1/N$ expansion should be zero. In order to turn a $1/N$ expansion into a {\it quantitative} tool, we need to:
1) understand the large order behavior of the series, 
2) locate the singularities of the Borel transform and, 
3) compare the accuracy of various procedures with numerical results for given values of $N$.
Calculating the series or obtaining accurate numerical results at fixed $N$ are difficult tasks and we do not know any model where this program has been completed. For instance for the critical exponents in three dimensions, we are only aware of calculation up to order $1/N^2$ in Ref. \cite{okabe78,gracey91,pelissetto00b}. 
Several results related to the possibility (or impossibility) of resumming particular $1/N$ expansions are known \cite{dewit77,avan83,kneur01}. Overall, it seems that there is a rather 
pessimistic impression regarding the possibility of using the $1/N$ expansion for low values of 
$N$. For this reason, it would be interesting to discuss the three questions enumerated above 
for a model where we have good chances to obtain definite answers. Dyson's hierarchical model \cite{dyson69,baker72} is a good candidate for this purpose.

In this Brief Report, we provide high-accuracy numerical values for the critical exponent $\gamma$, the subleading exponent $\Delta$ and the critical parameter $\beta_c$ for the $3D$ $O(N)$ hierarchical nonlinear sigma models.  These quantities appear in the 
magnetic susceptibility 
near $\beta_c$ in the symmetric phase as 
\begin{equation}
\chi= (\beta _c -\beta )^{-\gamma } (A_0 + A_1 (\beta _c -\beta)^{
\Delta }+\dots )\ .  \label{eq:sus}
\end{equation}

The method of calculation of the critical exponents used here is an 
extension of one of the 
methods described at length in the case of $N=1$ \cite{gam3} and will only be sketched briefly.  On the other hand, the accuracy of the approximations 
used depend non trivially on $N$ as we shall discuss later.
The RG transformation can be constructed as a blockspin transformation followed by a rescaling of the field. For Dyson's hierarchical model, the block spin transformation 
affects only the local measure. The RG transformation can be expressed conveniently in terms of 
the Fourier transform (denoted $R$ hereafter) of this local measure. In the following, we keep the $O(N)$ symmetry unbroken and the Fourier transform will depend only on $\vec{k} .\vec{k} \equiv u$. Here $\vec{k}$ is a source conjugated to the local field variable $\vec{\phi}$. Replacing $k$ by $u$ and the second derivative by the $N$-dimensional Laplacian in Eq. (2.5) of Ref. \cite{gam3}, we obtain the RG transformation for the Fourier transform of the local measure:
\begin{equation}
R_{n+1,N}(u)\propto 
{\rm e}^{\left[ -\frac {1}{2} \beta \left( 4u
\frac{\partial^2 }{\partial u^2}+
2N \frac{\partial }{\partial u} \right)
\right]}\left( R_{n,N}\left( c u/4 \right) \right)
^2 \ , \label{eq:recursionN}
\end{equation}
where $c=2^{1-2/D}$ in order to reproduce the scaling of a Gaussian massless field in $D$ 
dimensions. $D=3$ hereafter.
We fix the normalization constant by imposing 
$R_{n,N}(0)=1$ so that 
$R_{n,N}(k)$ has a simple probabilistic
interpretation \cite{gam3}. In the following, the calculations
will be performed using polynomial approximations of degree
$l_{\max}$:
\begin{equation}
R_{n,N}(k)\simeq 1+a_{n,1}u+a_{n,2}u^2+\cdots+a_{n,l_{\max}}u^{l_{\max}}\ .
\end{equation}
The finite volume susceptibility for $2^n$ sites is related to the first coefficient by the relation $\chi_n =-2a_{n,1}(2/c)^n$. 
The truncated recursion formula for the $a_{n,m}$ reads
\begin{equation}
\label{eq:quad}
a_{n+1,m}=\frac{\sum_{l=m}^{2l_{\max}}
\left(\sum_{p+q=l}a_{n,p}a_{n,q}\right)B_{m,l}}
{\sum_{l=0}^{2l_{\max}}\left(\sum_{p+q=l}a_{n,p}a_{n,q}\right)B_{0,l}}\ ,
\end{equation}
with 
\begin{equation}
B_{m,l}=\frac{\Gamma(l+1)\Gamma(l+N/2)}{\Gamma(m+1)\Gamma(m+N/2)}
\frac{1}{(l-m)!}(\frac{c}{4})^{l}(-2\beta)^{l-m}\ .
\end{equation}
We emphasize that in the above formula and in our numerical calculations, no truncation is applied after squaring and so the sum in Eq. (\ref{eq:quad}) does extend up to $2l_{max}$. 
Since the derivatives appear to arbitrarily large order in Eq. (\ref{eq:recursionN}) and can lower the degree of a polynomial of order larger than $l_{max}$, this 
affects all the coefficients of order less than $l_{max}$.
This procedure has been discussed and justified in Ref. \cite{scalingjsp}. 

The critical exponents appearing in Eq. (\ref{eq:sus}) are obtained by calculating the eigenvalues $\lambda_1, \lambda_2, \dots$ of the matrix $\partial a_{n+1,l}/\partial a_{n,m}$ at the nontrivial fixed point. The exponents 
$\gamma$ and $\Delta$, can be expressed as 
\begin{equation}
\label{eq:exponents}
\gamma=\frac{\ln(2/c)}{\ln(\lambda_1)} {\rm \ \ ,  \ \  }
\Delta=\left|\frac{\ln(\lambda_2)}{\ln(\lambda_1)}\right|\ .
\end{equation}

The critical exponents are universal and, within numerical errors, independent of the manner that we approach the nontrivial fixed point. In the following, we have mostly started with the 
local measure of the nonlinear sigma model $\delta (\vec{ \phi}. \vec{\phi}-1)$.  The corresponding Fourier transform reads 
\begin{equation}
R_{0,N}(u)=\sum_{l=0}^{\infty} \frac{(-1)^l
u^{l}\Gamma(\frac{N}{2})}{2^{2l}l!\Gamma(\frac{N}{2}+l)} \ .
\end{equation}
A motivation for this choice is that, as we will explain below, the value of $\beta_c$ can be calculated in the large $N$ limit. Other measures have also been used in order to 
check the universal values of the two exponents.

\begin{figure}[h]
\includegraphics[width=0.4\textwidth]{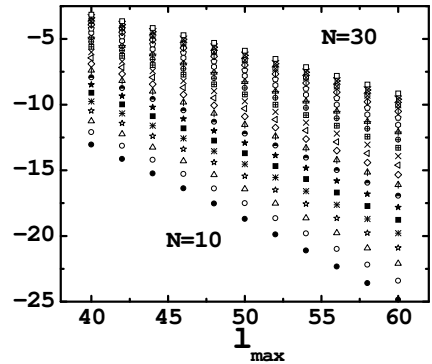}
\caption{$\log_{10}\frac
{|\beta(l_{\max})-\beta_c|}{\beta_c}$ calculated for 
$l_{\max}=40$ to $l_{\max}=60$ for $N=10$ (filled circles), $N=11$
(empty circles), $N=12$ (empty triangles) ...... up to $N=30$
(empty squares).\label{fig:errorlmax} } 
\end{figure}

\begin{figure}

\includegraphics[width=0.4\textwidth]{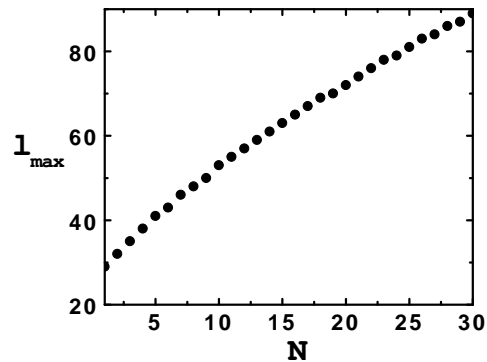}
\caption{Minimal value of $l_{\max}$ in order to have $\log_{10}\frac {|\beta_c(l_{\max})-\beta_c(\infty))|}{\beta_c(\infty)}=-20$ versus $N$.\label{fig:erro}} 
\end{figure}

The asymptotic behavior of the ratio
$a_{n+1,1}/a_{n,1}$  allows us to
decide unambiguously if we are in the symmetric phase (where the ratio approaches $c/2 \simeq 0.63$) or in the broken phase (where the ratio approaches $c $). Using a binary search, one can determine the critical value of $\beta$ with great accuracy. As this critical value depends on 
$l_{max}$, we denote it $\beta_c(l_{max})$. When $l_{max}\rightarrow\infty $, $\beta_c(l_{max})\rightarrow \beta_c$. The rate at which this limit is reached 
depends on $N$. This is illustrated in Fig. \ref{fig:errorlmax} where we see that in 
order to reach $\beta_c$ with a given accuracy, we need to increase $l_{max}$ when $N$ 
increases. In Fig. \ref{fig:erro}, we give the minimum $l_{max}$ necessary for $\beta_c(l_{max})$ to 
share 20 significant digits with $\beta_c$. $l_{max}\simeq 22+6.2N^{0.7}$ is a
good fit for Fig. \ref{fig:erro}. 

The nontrivial fixed point {\it for a given value of $l_{max}$} can be constructed  by iterating 
sufficiently many times the RG map at values sufficiently close to $\beta_c(l_{\max})$. 
In order to get an accuracy $\epsilon$ for the fixed point for that value of $l_{max}$, 
we need to iterate $n$ times the map until 
\begin{equation}
\lambda_2^n\sim \epsilon \ ,
\label{eq:irr}
\end{equation} in order to get rid 
of the irrelevant directions. 
At the same time, we want the growth in the relevant direction to be limited, in other words, 
\begin{equation}
|\beta -\beta_c(l_{\max})|\lambda _1 ^n <\epsilon\ .
\end{equation}
Combining these two requirements together with Eq. (\ref{eq:exponents}) we obtain 
\begin{equation}
|\beta -\beta_c(l_{\max})|\simeq \epsilon ^{1+1/\Delta}	
\end{equation}
This is an order magnitude estimate, however it works well except for $N$=1 where we need 
to pick $\beta$ slightly closer to the critical value. 
By ``working well'', we mean that if we go closer to the critical value, changes smaller than 
$\epsilon$ are observed in the first two eigenvalues.
The numerical results for $\epsilon =10^{-10}$ and $N$ up to 20, are given in the Tables \ref{table:one} and \ref{table:two} for the values of $l_{max}$ of Fig. \ref{fig:erro}.
Errors of 1 or less in the last printed digit should be understood in all the tables.
\begin{table}
\caption{$\beta_c $ and the first two eigenvalues for $N=1\dots 20$. \label{table:one}}
\begin{tabular}{||c|c|c|c||} 
\hline
$N$ & $\beta_c$ & $\lambda_1$ & $\lambda_2$ \cr
\hline
1 & 1.1790301704462697325 & 1.427172478 & 0.8594116492\cr 2 & 2.4735265752919854000 & 1.385743490 & 
   0.8563409066 \cr 3 & 3.8273820333573397671 & 1.354668326 & 0.8506945150 \cr 4 & 5.2111615635533656165 & 1.332749866 & 
   0.8440522956 \cr 5 & 6.6104153462855068435 & 1.317578283 & 0.8376436747 \cr 6 & 8.0181114053706725941 & 1.306955396 & 
   0.8320345022 \cr 7 & 9.4307096447427796882 & 1.299321025 & 0.8273378172 \cr 8 & 10.846330737925124699 & 1.293666393 & 
   0.8234676785 \cr 9 & 12.263918029354988652 & 1.289354227 & 0.8202833449 \cr 10 & 13.682844072802585664 & 1.285978489 & 
   0.8176485461 \cr 11 & 15.102717572108367579 & 1.283274741 & 0.8154492652 \cr 12 & 16.523283812777939366 & 1.281066141 & 
   0.8135953137 \cr 13 & 17.944370719047342283 & 1.279231192 & 0.8120168555 \cr 14 & 19.365858255947423937 & 1.277684252 & 
   0.8106600963 \cr 15 & 20.787660334686062513 & 1.276363511 & 0.8094834857 \cr 16 & 22.209713705054412233 & 1.275223389 & 
   0.8084547150 \cr 17 & 23.631970906283518487 & 1.274229622 & 0.8075484440 \cr 18 & 25.054395659078177206 & 1.273356000 & 
   0.8067446107 \cr 19 & 26.476959772907788848 & 1.272582158 & 0.8060271793 \cr 20 & 27.899641020779716433 & 1.271892050 & 
   0.8053832116\cr 
\hline
\end{tabular}
\end{table}

\begin{table}[h]
\caption{$\gamma$, $\Delta$ and $\beta_c/N$ for $N=1\dots 20$. \label{table:two}}
\begin{tabular}{||c|c|c|c||} 
\hline
$N$& $\gamma$ & $\Delta$ & $\beta_c/N$ \cr
\hline
1 & 1.29914073 & 0.425946859 & 1.179030170 \cr 2 & 1.41644996 & 0.475380831 & 1.236763288 \cr 3 & 
   1.52227970 & 0.532691965 & 1.275794011 \cr 4 & 1.60872817 & 0.590232008 & 1.302790391 \cr 5 & 1.67551051 & 
   0.642369187 & 1.322083069 \cr 6 & 1.72617703 & 0.686892637 & 1.336351901 \cr 7 & 1.76479863 & 0.723880426 & 
   1.347244235 \cr 8 & 1.79469274 & 0.754352622 & 1.355791342 \cr 9 & 1.81827105 & 0.779508505 & 1.362657559 \cr 10 & 
   1.83722291 & 0.800424484 & 1.368284407 \cr 11 & 1.85272636 & 0.817977695 & 1.372974325 \cr 12 & 1.86561092 & 
   0.832855522 & 1.376940318 \cr 13 & 1.87646998 & 0.845589221 & 1.380336209 \cr 14 & 1.88573562 & 0.856588705 & 
   1.383275590 \cr 15 & 1.89372812 & 0.866171682 & 1.385844022 \cr 16 & 1.90068903 & 0.874586271 & 
   1.388107107 \cr 17 & 1.90680338 & 0.882027998 & 1.390115936 \cr 18 & 1.91221507 & 0.888652409 & 
   1.391910870 \cr 19 & 1.91703752 & 0.894584429& 1.393524199 \cr 20 & 1.92136121 & 0.899925325 & 1.394982051 \cr 
   $\infty$ & 2& 1& $\frac{2-c}{2(c-1)}=1.42366..$ \cr
\hline
\end{tabular}
\end{table}

As $N$ increases, the values displayed in Table \ref{table:two} seem to slowly approach asymptotic values. This is expected. Using the general formulation of Ref. \cite{ma73,david85} together with the particular form of the propagator \cite{complexs} for the model considered here, one finds the leading terms
\begin{eqnarray}
\label{eq:next}
\gamma &\simeq & 2+a_1/N+\dots \\ \nonumber 
\Delta &\simeq & 1+b_1/N +\dots \\ 
\beta_c/N &\simeq & (2-c)/(2(c-1)) +c_1/N +\dots \ .
\end{eqnarray}
The magnitude of the coefficients $a_1, \ b_1,\ c_1$ of the leading $1/N$ corrections can be estimated by subtracting the asymptotic value and multiplying by $N$. The results are shown in  
Table \ref{table:three}. They indicate that $a_1 \simeq  -1.6, \ b_1 \simeq -2.0,\ c_1 \simeq -0.57$. It seems possible to improve the accuracy by estimating the next to leading order corrections and so on. However, the stability of this procedure is more delicate and 
remains to be studied with simpler examples.
\begin{table}
\vskip15pt
\caption{$N(2-\gamma )$, $N(1-\Delta)$ and $N(\frac{2-c}{2(c-1)}-\frac{\beta_c}{N})$  
for $N=1\dots 20$.. \label{table:three}}
\vskip15pt
\begin{tabular}{||c|c|c|c||} 
\hline
$N$ & $N(2-\gamma )$ & $N(1-\Delta)$ & $N(\frac{2-c}{2(c-1)}-\frac{\beta_c}{N})$\cr
\hline
1 & 0.7009 & 0.5741 & 0.2446 \cr 2 & 1.167 & 1.049 & 0.3738 \cr 3 & 1.433 & 1.402 & 
   0.4436 \cr 4 & 1.565 & 1.639 & 0.4835 \cr 5 & 1.622 & 1.788 & 0.5079 \cr 6 & 1.643 & 1.879 & 
   0.5239 \cr 7 & 1.646 & 1.933 & 0.5349 \cr 8 & 1.642 & 1.965 & 0.5430 \cr 9 & 1.636 & 1.984 & 
   0.5490 \cr 10 & 1.628 & 1.996 & 0.5538 \cr 11 & 1.620 & 2.002 & 0.5576 \cr 12 & 1.613 & 2.006 & 
   0.5606 \cr 13 & 1.606 & 2.007 & 0.5632 \cr 14 & 1.600 & 2.008 & 0.5654 \cr 15 & 1.594 & 2.007 & 
   0.5673 \cr 16 & 1.589 & 2.007 & 0.5689 \cr 17 & 1.584 & 2.006 & 0.5703 \cr 18 & 1.580 & 2.004 & 
   0.5715 \cr 19 & 1.576 & 2.003 & 0.5726 \cr 20 & 1.573 & 2.001 & 0.5736 \cr 
\hline
\end{tabular}
\end{table}

We now compare the exponents calculated here with those calculated with three other RG transformations \cite{comellas97,gottker99,litim02}. As we proceed to explain, the exponents should be the same in the four cases (including ours). The change of coordinates that relates the RG transformation considered here and the one studied in Ref. \cite{gottker99} is given in the introduction of \cite{koch91} (for $L=2^{1/3}$). 
The fact that the limit $L \rightarrow 1$ in the formulation of Ref. \cite{gottker99} yields 
the Polchinski equation in the local potential approximation studied in Ref. \cite{comellas97} is explained in Ref. \cite{felder87}.
Consequently, these two RG transformations should be the same in the {\it linear} approximation.
Finally, Litim \cite{litim00,litim02} proposed an optimized version of the exact RG transformation and 
suggested \cite{litim05} that it was equivalent to the Polchinski equation in the local potential approximation. The equivalence was subsequently proved by Morris \cite{morris05}.

To facilitate the comparison, we display $\nu =\gamma /2$ (since $\eta =0$ here) and  $\omega = \Delta/\nu $ in Table \ref{table:four}. Our results coincide with the 4 digits given in 
column (2) of Table 3 (for $\nu$) and 4 (for $\omega$) in \cite{comellas97}. They coincide with the six digits for $\nu$ given in the line $d=3$ of Table 8 of \cite{gottker99} for $N$= 1, 2, 3, 5 and 10. However, we found discrepancies of order 1 in the fifth digit of $\nu$ and slightly 
larger for $\omega$ with the values found in Table 1  of  \cite{litim02}. Our estimated errors are of order 1 in the 9-th digit. For $N=1$, this is confirmed by an independent method \cite{gam3}. For $N=$ 2, 3, 5, and 10, this is confirmed up to the sixth digit \cite{gottker99}. Consequently, 
a discrepancy in the 5-th digit cannot be explained by our numerical errors. Note also that for  $N\geq 2$, $\alpha$ is more negative than for nearest neighbor models \cite{pelissetto00b}.

In summary, we have provided high-accuracy data for $\gamma$, $\Delta$ and $\beta_c$
 for $N$ up to 20. It seems likely that a few terms of the $1/N$ expansion for these three quantities can be estimated from this data. Work is in progress to calculate these 
expansions independently by semi-analytical methods and learn about the asymptotic behavior of 
the series and their accuracy. The discrepancy with the 5-th digit of Ref. \cite{litim02} remains to be explained. 

We thank G. 't Hooft, J. Zinn-Justin and G. Parisi for valuable comments on related topics. 
This research was supported in part by the Department of Energy
under Contract No. FG02-91ER40664.
M.B. Oktay has been supported by SFI grant
          04/BRG/P0275.

\begin{table}[t]
\caption{$\nu$, $\omega$ and $\alpha$ for $N=1\dots 20$. \label{table:four}}
\vskip15pt
\begin{tabular}{||c|c|c|c||} 
\hline
$N$ & $\nu =\gamma /2$ & $\omega = \Delta/\nu $& $\alpha=2-3\nu$\cr
\hline
 1 & 0.649570 & 0.655736  & 0.051289  \cr 
 2 & 0.708225 & 0.671229  & -0.124675   \cr 
 3 & 0.761140& 0.699861   &  -0.283420   \cr 
 4 & 0.804364 & 0.733787  &   -0.413092 \cr 
 5 & 0.837755 & 0.766774  &   -0.513266\cr 
 6 & 0.863089 & 0.795854  &  -0.589266\cr 
 7 & 0.882399 & 0.820355  &  -0.647198\cr 
 8 & 0.897346 & 0.840648  &  -0.692039\cr 
 9 & 0.909136 & 0.857417  &  -0.727407\cr 
 10 & 0.918611 & 0.871342 &    -0.755834\cr 
\hline
\end{tabular}
\end{table}

\end{document}